\documentclass{wqcd03}                

\confname{QCD@Work 2003 - International Workshop on QCD,
Conversano, Italy, 14--18 June 2003}

\title{Connecting Polyakov Loops to Hadrons}
\author{A.~Mocsy\addressmark{a}, F.~Sannino\addressmark{a}\thanks{Speaker at the Workshop} and
K.~Tuominen \addressmark{a} }

\address[a]{NORDITA and The Niels Bohr Institute,  Blegdamsvej 17, DK-2100 Copenhagen \O, Denmark }

\begin{document}

\begin{abstract}
The order parameter for the pure Yang-Mills phase transition is
the Polyakov loop, which encodes the symmetries of the $Z_N$
center of the $SU(N)$ gauge group. The physical degrees of freedom
of any asymptotically free gauge theory are hadronic states. Using
the Yang-Mills trace anomaly and the exact $Z_N$ symmetry we show
that the transfer of information from the order parameter to
hadrons is complete.
\end{abstract}

\maketitle

\section{Introduction}
\label{uno}

Recently in
\cite{Sannino:2002wb,{Mocsy:2003tr},{Mocsy:2003un},{Mocsy:2003qw}}
we have analyzed the problem of how, and to what extent the
information encoded in the order parameter of a generic theory is
transferred to the non-order parameter fields. This is a
fundamental problem since in nature most fields are non-order
parameter fields.

This problem is especially relevant in QCD and QCD-like theories
since there is no physical observable for deconfinement which is
directly linked to the order parameter field. Here we mainly
review the first work \cite{Sannino:2002wb} on this issue, which
deals with the confinement/deconfinement phase transition in pure
Yang-Mills theory.

Importance sampling lattice simulations are able to provide vital
information about the nature of the temperature driven phase
transition for 2 and 3 color Yang-Mills theories with and without
matter fields (see \cite{Boyd:1996bx,{Okamoto:1999hi}} for 3
colors). At zero temperature the $SU(N)$ Yang-Mills theory is
asymptotically free, and the physical spectrum of the theory
consists of a tower of hadronic states referred to as glueballs
and pseudo-scalar glueballs. The theory develops a mass gap and
the lightest glueball has a mass of the order of few times the
confining scale. The classical theory possesses conformal
symmetry, while quantum corrections lead to a non-vanishing trace
of the energy momentum tensor.

At nonzero temperature the $Z_N$ center of $SU(N)$ is a relevant
global symmetry \cite{Svetitsky:1982gs}, and it is possible to
construct a number of gauge invariant operators charged under
$Z_N$. Among these the most notable one is the Polyakov loop:
\begin{eqnarray} {\ell}\left(x\right)=\frac{1}{N}{\rm Tr}[{\bf L}]\equiv\frac{1}{N}{\rm Tr}
\left[{\cal
P}\exp\left[i\,g\int_{0}^{1/T}A_{0}(x,\tau)d\tau\right]\right]
\nonumber
\end{eqnarray} ${\cal P}$ denotes path ordering, $g$ is the $SU(N)$ coupling constant,
$x$ is the coordinate for the three spatial dimensions, and $\tau$
is the Euclidean time. The $\ell$ field is real for $N=2$, while
otherwise complex. This object is charged with respect to the
center $Z_N$ of the $SU(N)$ gauge group \cite{Svetitsky:1982gs},
under which it transforms as $\ell \rightarrow z \ell$ with $z\in
Z_N$. A relevant feature of the Polyakov loop is that its
expectation value vanishes in the low temperature regime, and is
non-zero in the high temperature phase. The Polyakov loop is thus
a suitable order parameter for the Yang-Mills temperature driven
phase transition \cite{Svetitsky:1982gs}.

Here we consider pure gluon dynamics. This allows us to have a
well defined framework where the $Z_N$ symmetry is exact. The
hadronic states are the glueball fields ($H$) and their effective
theory at the tree level is constrained by the Yang-Mills trace
anomaly.

A puzzle is how the information about the Yang-Mills phase
transition encoded, for example, in the $Z_N$ global symmetry can
be communicated to the hadronic states of the theory. Here we
propose a concrete way to resolve the puzzle.

As basic ingredients we use the trace anomaly and the $Z_N$
symmetry. They will be enough to demonstrate that the information
carried by $\ell$ is efficiently and completely transferred to the
glueballs. More generally, the glueball field is a function of
$\ell$:
\begin{eqnarray} H\equiv H[\ell] . \end{eqnarray} Our results
can be tested via first principle lattice simulations.

\section{The Basic Properties}
\label{due} The hadronic states of the Yang-Mills theory are the
glueballs.  At zero temperature the Yang-Mills trace anomaly has
been used to constrain the potential of the lightest glueball
state $H$ \cite{Schechter:2001ts}:
\begin{eqnarray}
V[H]=\frac{H}{2} \ln \left[\frac{H}{\Lambda^4}\right] .
\end{eqnarray}
$\Lambda$ is chosen to be the confining scale of the theory and
$H$ is a mass dimension four field. This potential correctly
saturates the trace anomaly when $H$ is assumed to be proportional
to ${\rm Tr}\left[G_{\mu \nu}G^{\mu \nu}\right]$ and $G_{\mu \nu}$
is the standard Yang-Mills field strength. The potential nicely
encodes the properties of the Yang-Mills vacuum at zero
temperature and it has been used to deduce a number of
phenomenological results \cite{Schechter:2001ts}. Effective
Lagrangians play a relevant role for describing strong dynamics.
Recently, for example a number of fundamental results at zero
temperature and matter density about the vacuum properties and
spectrum of QCD have been uncovered \cite{Sannino:2003xe}.

At high temperature the Yang-Mills pressure can be written in
terms of the field $\ell$ according to the Polyakov Loop Model
(PLM) of \cite{Pisarski:2001pe}. This free energy must be
invariant under $Z_N$ and it takes the general form:
\begin{eqnarray}
V[\ell]= T^4{\cal F}[\ell] .
\end{eqnarray}
${\cal F}[\ell]$ is a polynomial in $\ell$ invariant under $Z_N$,
and its coefficients depend on the temperature, allowing for a
mean field description of the Yang-Mills phase transitions.

We now marry the two models by requiring both fields to be present
simultaneously at non zero temperature. The theory must reproduce
the ordinary glueball Lagrangian at low temperatures and the PLM
Lagrangian at high temperatures. In \cite{Sannino:2002wb}
 the following effective potential was proposed:
\begin{equation}
V\left[H,\ell \right]
=\frac{H}{2}\ln\left[\frac{H}{\Lambda^4}\right] +
V_{T}\left[H\right]+H{\cal P}{\left[ \ell \right]} + T^4 {\cal
V}\left[ \ell \right]   , \label{POTENTIAL}
\end{equation}
where ${\cal V}\left[ \ell \right]$ and ${\cal P}\left[ \ell
\right]$ are general (but real) polynomials in $\ell$ invariant
under $Z_N$ whose coefficients depend on the temperature. The
explicit dependence is not known and should be fit to lattice
data. Dimensional analysis and analyticity in $H$ when coupling it
with $\ell$ severely restricts the effective potential terms. We
stress that $H{\cal P}[\ell]$ is the most general interaction term
which can be constructed without spoiling the zero temperature
trace anomaly. Further nonanalytic interaction terms can arise
when considering thermal and quantum corrections. These are
partially contained in $V_{T}[H]$ which schematically represents
the temperature of a gas of glueballs. In the following we will
not investigate in detail such a term. Our theory cannot be the
full story since we neglected (as customary) all of the tower of
glueballs and pseudo-scalar glueballs as well as the infinite
series of dimensionless gauge invariant operators with different
charges with respect to $Z_N$. Nevertheless, the potential is
sufficiently general to capture the essential features of the
Yang-Mills phase transition.

When the temperature $T$ is much less than the confining scale
$\Lambda$ the last term in Eq.~(\ref{POTENTIAL}) can be safely
neglected. Since the glueballs are relatively heavy compared to
the $\Lambda$ scale their temperature contribution $V_T[H]$ can
also be disregarded. At low temperatures the theory reduces to the
standard glueball potential augmented by the third term which does
not affect the trace anomaly.

At very high temperatures (compared to $\Lambda$) the last term
dominates ($H$ itself is very small) recovering the picture in
which $\ell$ dominates the free energy. In this regime we have
${\cal F}[\ell]={\cal V}[\ell]$.

We can, in principle, compute all the relevant thermodynamic
quantities in our approach, i.e. entropy, pressure etc., but this
is not the main scope of this work.

A relevant object is the trace of the energy-momentum tensor
$\Theta^{\mu}_{\mu}$. At zero temperature, and when the potential
is a general function of a set of bosonic fields $\{\Phi_n \}$
with mass-dimensions $d_n$, one can construct the associated trace
of the energy-momentum tensor  via:
\begin{equation}
\Theta^{\mu}_{\mu}=4V[\Phi_n]-\sum_n\frac{\delta V[\Phi_n]}{\delta
\Phi_n}\Phi_n\, d_n .  \label{Ttheta}
\end{equation}
At finite temperature we still define our temperature dependent
energy-momentum tensor as in Eq.~(\ref{Ttheta}). Here $H$
possesses engineering mass dimensions $4$ while  $\ell$  is
dimensionless, yielding the following temperature dependent stress
energy tensor:
\begin{equation}
\Theta^{\mu}_{\mu}(T)= -2H +4T^4{\cal V}[\ell] +
4\left[1-H\frac{\delta}{\delta H}\right]V_T [H] .
\end{equation}
$\Theta^{\mu}_{\mu}$ is normalized such that $ \langle 0|
\Theta^{\mu}_{\mu} |0\rangle = \epsilon - 3p $ with $\epsilon$ the
vacuum energy density and $p$ the pressure. At zero temperature
only the first term survives, yielding magnetic type condensation
typical of a confining phase, while at extremely high temperature
the second term dominates, displaying an energy density and a
pressure typical of a deconfined phase.

The theory containing just $\ell$  can be obtained integrating out
$H$ via the equation of motion:
\begin{eqnarray}
\frac{\delta V[H,\ell]}{\delta H}= 0\, .   \label{OUT}
\end{eqnarray}
{}Formally this is justifiable if the glueball degrees of freedom
are very heavy. {}For simplicity we neglect the contribution of
$V_T[H]$, as well as the mean-field theory corrections for $\ell$.
However, a more careful treatment which also includes the kinetic
terms should be considered \cite{Mocsy:2003tr,MocsyProcs}. Within
these approximations the equation of motion yields:
\begin{eqnarray}
H[\ell]=\frac{\Lambda^4}{e}\exp\left[ -2 {\cal P}[\ell] \right] .
\label{Hofl}
\end{eqnarray}
The previous expression shows the intimate relation between $\ell$
and the physical states of strongly interacting theories.

After substituting Eq.~(\ref{Hofl}) back into the potential
(\ref{POTENTIAL}) and having neglected $V_T[H]$ we have:
\begin{eqnarray}
V[\ell]=T^4 {\cal V}[\ell] - \frac{\Lambda^4}{2e}\exp \left[-2
{\cal P}[\ell] \right] .
\end{eqnarray}
This formula shows that for large temperatures the only relevant
energy scale is $T$ and we recover the PLM model. At low
temperatures the scale $\Lambda$ allows for new terms in the
Lagrangian. Besides the $T^4$ and the $\Lambda^4$ terms we also
expect terms with coefficients of the type $T\Lambda^3$ and
$T^2\Lambda^2$ and $T^3 \Lambda$. However in our simple model
these terms do not seem to emerge.

Expanding the exponential we have:
\begin{eqnarray}
V[\ell]=T^4 {\cal V}[\ell] +\frac{\Lambda^4}{e}{\cal P}[\ell] -
\frac{\Lambda^4}{2 e} + \cdots .
\end{eqnarray}
Since ${\cal V}[\ell]$ and ${\cal P}[\ell]$ are real polynomials
in $\ell$ invariant under $Z_N$ we immediately recover a general
potential  in $\ell$.

\section{The two Color Theory}
\label{tre} To illustrate how our formalism works we first
consider in detail the case $N=2$ and neglect for simplicity the
term $V_T[H]$. This theory has been extensively studied via
lattice simulations \cite{Damgaard,{Hands:2001jn}} and it
constitutes the natural playground to test our model.  Here $\ell$
is a real field and the $Z_2$ invariant ${\cal V}[\ell]$ and
${\cal P}[\ell]$ are taken to be:
\begin{eqnarray}
   {\cal V}\left[ \ell \right] &=& a_1 \ell ^2 + a_2 \ell^4  + {\cal
   O}(\ell^6) , \nonumber \\
     {\cal P}\left[ \ell \right] &=& b_1 \ell ^2 + {\cal
   O}(\ell^4) ,
\end{eqnarray}
with $a_1,~a_2$ and $b_1$ unknown temperature dependent functions,
which should be derived directly from the underlying theory.
Lattice simulations can, in principle, fix all of the
coefficients. In order for us to investigate in some more detail
the features of our potential, and inspired by the PLM model
mean-field type of approximation, we first assume $a_2$ and $b_1$
to be positive and temperature independent constants, while we
model $a_1=\alpha (T_{\ast}-T)/T$,  with $T_{\ast}$ a constant and
$\alpha$ another positive constant. We will soon see that due to
the interplay between the hadronic states and $\ell$,  $T_{\ast}$
need not to be the critical Yang-Mills temperature while $a_1$
displays the typical behavior of the mass square term related to a
second order type of phase transition.

The extrema are obtained by differentiating the potential with
respect to $H$ and $\ell$:
\begin{eqnarray}
\frac{\partial V}{\partial H}&=&
\frac{\ln}{2}\left[\frac{eH}{\Lambda^4}\right]+{\cal
P}\left[\ell\right]
=\frac{\ln}{2}\left[\frac{eH}{\Lambda^4}\right]+b_1\ell^2=0
\label{motiona} \\ \frac{\partial V}{\partial \ell}&=&2\ell
T^4\left(a_1 + \frac{H}{T^4} b_1 + 2a_2 \ell^2 \right) =0 \, .
\label{motionb}
\end{eqnarray}

\subsection{Small and Intermediate Temperatures}
At small temperatures the second term in Eq.~(\ref{motionb})
dominates and the only solution is $\ell=0$. A vanishing $\ell$
leads to a null ${\cal P}[\ell]$ yielding  the expected minimum
for $H$:
\begin{equation}
\langle H\rangle = \frac{\Lambda^4}{e} .  \label{Hvacuum}
\end{equation}
Here $\ell$ and $H$ decouple.

We now study the solution near the critical temperature for the
deconfinement transition. {}At all temperatures for which
\begin{eqnarray}
T^4 a_1 + {H} b_1 =T^3 \alpha (T_{\ast}-T) + H b_1 > 0 ,
\end{eqnarray}
the solution for $\ell$ is still $\ell=0$, yielding
Eq.~(\ref{Hvacuum}). The critical temperature is reached for
\begin{eqnarray}
T_c = T_{\ast} + \frac{b_1}{e\alpha}\frac{\Lambda^4}{T_c^3} \, ,
\label{criticalTc}
\end{eqnarray}
and can be determined via lattice simulations. We see that within
our framework the latter is related to the glueball
(gluon-condensate) coupling to two Polyakov loops and it would be
relevant to measure it on the lattice.  At $T=T_c$, $\ell=0$ and
$H=\Lambda^4 /e$.

Let us now consider the case $T=T_c + \Delta T$ with
\begin{eqnarray}\frac{\Delta T}{T_c}  \ll 1 .\end{eqnarray}
Expanding $\langle \ell \rangle ^2$ at the leading order in
$\Delta T/T_c$ yields:
\begin{eqnarray}
\langle \ell \rangle^2 =\frac{\alpha}{2a_2}
\frac{1+3\frac{b_1}{e\alpha} \frac{\Lambda^4}{T_c^4}}{1 -
\frac{b_1^2}{ e a_2} \frac{\Lambda^4}{T_c^4}} \, \frac{\Delta
T}{T_c} .
\end{eqnarray}
We used Eq.~(\ref{criticalTc}) and Eq.~(\ref{motiona}) which
relates the temperature dependence of $H$ to the one of $\ell$. At
high temperatures (see next subsection) $\langle \ell \rangle $
can be normalized to one by imposing $\alpha/2a_2 =1$ and the
previous expression reads:
 \begin{eqnarray}
\langle \ell \rangle^2 = \frac{1+3\frac{b_1}{e\alpha}
\frac{\Lambda^4}{T_c^4}}{1 - \frac{2b_1^2}{e \alpha}
\frac{\Lambda^4}{T_c^4}} \, \frac{\Delta T}{T_c} \equiv\frac{4
T_c-3T_{\ast}}{(1-2b_1)T_c + 2b_1 T_{\ast}}\,\frac{\Delta T}{T_c}.
\end{eqnarray}
{}For a given critical temperature consistency requires $b_1$ and
$T_{\ast}$  to be such that:
\begin{eqnarray} \frac{4 T_c-3T_{\ast}}{(1-2b_1)T_c + 2b_1 T_{\ast}}
\geq 0 .
\end{eqnarray}  In this regime, the temperature dependence of the gluon
condensate is:
\begin{eqnarray}
\langle H \rangle =\frac{\Lambda^4}{e}\, \exp\left[-2b_1 \langle
\ell \rangle ^2 \right]  .
\end{eqnarray}
We find the mean field exponent for $\ell$, i.e. $\ell^2$
increases linearly with the temperature near the phase transition
\footnote{Corrections to the mean field are large and must be
taken into account.}. Interestingly, the gluon-condensates drops
exponentially. This drop is triggered by the rise of $\ell$ and it
happens in our simple model exactly at the deconfining critical
temperature. Although the drop might be sharp it is continuous in
temperature. This is related to the fact that the phase transition
is second order. Our findings strongly support the common picture
according to which the drop of the gluon condensate signals, in
absence of quarks, deconfinement.

\subsection{High Temperature}
\begin{figure}[h]
\begin{center}
\includegraphics[width=7truecm]{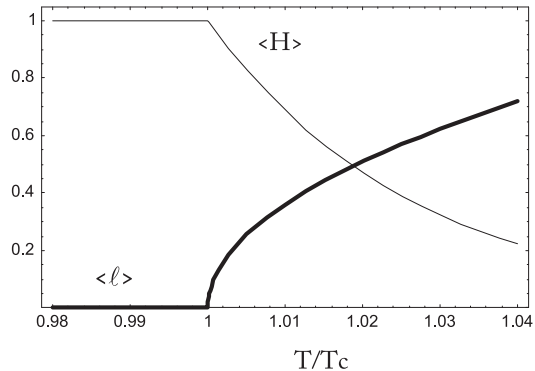}
\end{center}
\caption{The thin line is the gluon condensate $\langle H \rangle$
normalized to $\Lambda^4/e$ as function of temperature. The thick
line is $\langle \ell \rangle $ normalized to unity as function of
temperature. We chose for illustration $\alpha=1$, $b_1=1.45$ and
$T_c\simeq 1.16 \Lambda$.} \label{Figura1}
\end{figure}

 At very high temperatures the second
term in Eq.~(\ref{motionb}) can be neglected and the minimum for
$\ell$ is:
\begin{equation}
\langle \ell \rangle =\sqrt{\frac{\alpha}{2a_2}}. \label{HighT}
\end{equation}
{}For $H$ we have now:
\begin{eqnarray}
\langle H \rangle = \frac{\Lambda^4}{e} \exp \left[
-2b_1\frac{\alpha }{2 a_2} \right] = \frac{\Lambda^4}{e} \exp
\left[ - 2b_1 \right]  .
\end{eqnarray}
In the last step we normalized $\langle \ell \rangle$ to unity at
high temperature. In order for the previous solutions to be valid
we need to operate in the following temperature regime:
\begin{eqnarray}
T \gg \sqrt[4]{\frac{b_1}{\alpha} {\langle H \rangle}}\approx T_c.
\end{eqnarray}
  We find that at sufficiently high temperature $\langle H \rangle$ is exponentially
suppressed and the suppression rate is determined solely by the
glueball -- $\ell^2$ mixing term encoded in ${\cal P}[\ell]$. The
coefficient $b_1$ should be large (or increase with the
temperature) since we expect a vanishing gluon-condensate at
asymptotically high temperatures. Clearly it is crucial to
determine all of these coefficients via first principle lattice
simulations. The qualitative picture which emerges in our analysis
is summarized in Fig.~\ref{Figura1}.

\section{The three color theory}
\label{quattro} 
\begin{figure}[h]
\begin{center}
\includegraphics[width=7truecm]{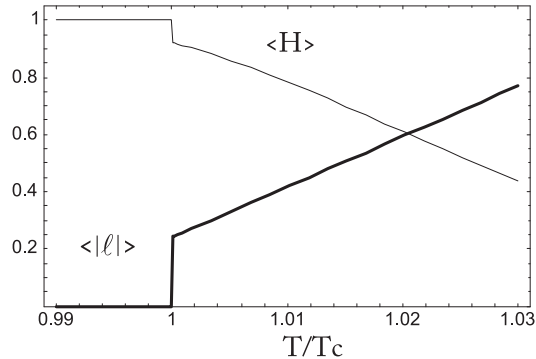}
\end{center}
\caption{A cartoon sketching the gluon condensate $\langle H
\rangle$ normalized to $\Lambda^4/e$ and the $\langle |\ell|
\rangle $ (thick line) as function of temperature. We chose for
illustration $a_3=0.3$,  $a_2=1$, $\alpha=2$, $b_1=0.7$ and
$T_c=1.2 \Lambda$.} \label{Figura2}
\end{figure}

$Z_3$ is the global symmetry group for the three
color case and $\ell$ is a complex field. The functions ${\cal
V}[\ell]$ and ${\cal P}[\ell]$ are:
\begin{eqnarray}
     {\cal V}\left[ \ell \right] &=& a_1 |\ell| ^2 +
     a_2 |\ell|^4  -a_3 (\ell^3 + {\ell^{\ast}}^3)+{\cal
   O}(\ell^5)  ,\nonumber \\
     {\cal P}\left[ \ell \right] &=& b_1 |\ell| ^2 + {\cal
   O}(\ell^3)  ,
   \label{trepotenziali}
\end{eqnarray}
with $a_1$, $a_2$, $a_3$ and $b_1$ unknown temperature dependent
coefficients which can be determined using lattice data. Here we
want to investigate the general relation between glueballs and
$\ell$, so we will not try to find the best parameterization to
fit the lattice data. In the spirit of the mean field theory we
take $a_2$, $a_3$ and $b_1$ to be positive constants while
$a_1=\alpha (T_{\ast}-T)/T$. With $\ell = |\ell| e^{i\varphi}$ the
extrema are now obtained by differentiating the potential with
respect to $H$, $|\ell|$ and $\varphi$:
\begin{eqnarray}
\frac{\partial V}{\partial H}&=&
\frac{\ln}{2}\left[\frac{eH}{\Lambda^4}\right]+{\cal
P}\left[\ell\right]
=\frac{\ln}{2}\left[\frac{eH}{\Lambda^4}\right]+b_1|\ell|^2=0
\nonumber  \\ \frac{\partial V}{\partial |\ell|}&=&2|\ell|
T^4\left(a_1 + \frac{H}{T^4} b_1 -3a_3 |\ell|\cos(3\varphi)  +
2a_2 |\ell|^2 \right) =0 \nonumber \\\frac{\partial V}{\partial
\varphi}&=& 6|\ell|^3 \, \sin(3\varphi)= 0\, .  \label{motion3}
\end{eqnarray}
At small temperature the $H/T^4$ term in the second equation
dominates and the solution is $\langle |\ell| \rangle =0$,
$\langle H \rangle=\Lambda^4/e$. The last equation is verified for
any $\langle \varphi \rangle$, so we choose $\langle \varphi
\rangle = 0$. The second equation can have two more solutions:
\begin{eqnarray}
 \frac{3}{4}\frac{a_3}{a_2} \pm
\sqrt{\frac{9}{16}\frac{a_3^2}{a_2^2} +\frac{\alpha
(T-T_{\ast})}{2 T a_2} - \frac{b_1H}{2a_2 T^4} }  ,
\end{eqnarray}
whenever the square root is well defined, i.e. at sufficiently
high temperatures. The negative sign corresponds to a relative
maximum, while the positive one to a relative minimum. We have
then to evaluate the free energy value (i.e. the effective thermal
potential) at the relative minimum and compare it with the one at
$\ell=0$. The temperature value for which the two minima have the
same free energy is defined as the critical temperature and is:
\begin{eqnarray}
T_c = \left[T_{\ast} + \frac{b_1}{e\alpha}\frac{\Lambda^4}{T_c^3}
\right] \frac{\alpha a_2}{\alpha a_2 + a_3^2}  .
\label{criticalTc3}
\end{eqnarray}
{}When $a_3$ vanishes we recover the second order type critical
temperature. To derive the previous expression we held the value
of $H$ fix to $\Lambda^4/e$ at the transition point. In a more
refined treatment one should not make such an assumption.
 Below this temperature the minimum is still for
$\langle \ell\rangle=0$ and $\langle H \rangle =\Lambda^4 /e$.
Just above the critical temperature the fields jump to the new
values:
\begin{eqnarray}
\langle |\ell|\rangle = \frac{a_3}{a_2} , \qquad \langle H \rangle
=\frac{\Lambda^4}{e} \exp{\left[-2 b_1 \langle |\ell| \rangle^2
\right]}  .
\end{eqnarray}
Close to, but above $T_c$ (i.e. $T=T_c +\Delta T$) we have:
\begin{eqnarray}
\langle |\ell| \rangle \simeq \frac{a_3}{a_2}+\rho \frac{\Delta
T}{T_c}
 ,
\end{eqnarray}
with \begin{eqnarray} \rho & \simeq& \frac{\alpha a_2}{a_3}
\frac{4 \kappa T_c -3 T_{\ast} }{a_2 T_c -4 b_1 \alpha (\kappa T_c
-
T_{\ast})} , \nonumber \\ \kappa &=& \frac{\alpha a_2 +
a_3^2}{\alpha a_2} \, ,
\end{eqnarray} a positive function of the coefficients of the
effective potential. In this regime
\begin{eqnarray}
\langle H \rangle = \frac{\Lambda^4}{e} \exp\left[-2
b_1(\frac{a_3}{a_2} + \rho \frac{\Delta T}{T_c})^2 \right]  .
\end{eqnarray}
At high temperature we expect a behavior similar to the one
presented for the two color theory. A cartoon representing the
behavior of the Polyakov loop and the gluon condensate is
presented in Fig.~\ref{Figura2}.

Since we are in the presence of a first order phase transition
higher order terms in Eq.~(\ref{trepotenziali}) may be important.
Lattice simulations have shown however, that the behavior of the
Polyakov loop for 3 colors resemble a weak first order transition
(i.e. small $a_3$) partially justifying our simple approach. The
approximation for our coefficients is too crude and it would
certainly be relevant to fit them to lattice simulations.

What we have learned is that the gluon condensate, although not a
real order parameter, encodes the information of the underlying
$Z_3$ symmetry. More generally, we have shown that once the map
between hadronic states and the true order parameter is known, we
can use directly hadronic states to determine when the phase
transition takes place and also determine the order of the phase
transition. {}For example by comparing Fig.~{\ref{Figura1}} and
Fig.~\ref{Figura2} we immediately notice the distinct behaviors in
the temperature dependence of the gluon condensate near the phase
transition.

\section{Conclusions}
\label{cinque} Our new theory is able to account for many features
inherent to the Yang-Mills deconfining phase transition. We
related two very distinct and relevant sectors of the theory: the
hadronic sector (the glueballs), and some dimensionless fields
($\ell$) charged under the discrete group $Z_N$ understood as the
center of the underlying $SU(N)$ Yang-Mills theory.

The gluon-condensate is, strictly speaking, not an order parameter
for the deconfining Yang-Mills phase transition. However we have
shown that the information encoded in the true order parameter
$\ell$ is efficiently communicated to the gluon condensate. Since
the exponential drop of the condensate just above the Yang-Mills
critical temperature is a direct consequence of the behavior of
the true order parameter at the transition we can consider this
drop as a strong signal of deconfinement. This drop has already
been used in various models for the Yang-Mills phase transition.
We have also seen that the reduction in the gluon-condensate is
associated to the increase of the Polyakov loop condensate $\ell$.
The information about the order of the phase transition is also
transferred to the behavior of the gluon condensate. Indeed, from
Fig.~\ref{Figura1} and Fig.~\ref{Figura2} we deduce that the drop
is continuous for the gluon condensate in the two color case,
while is discontinuous for the three color theory. We now have a
deeper understanding of the mechanism for transferring information
from the Yang-Mills order parameter to the physical states. Other
theoretical investigations also seem to support the present
results \cite{Meisinger:2002ji}.

It is worth mentioning that the Polyakov loop need not to be the
only acceptable order parameter. For example using an abelian
projection one can define a new  (non local in the cromomagnetic
variables) order parameter \cite{Giacomo:2002qr}. Our model can
be, in principle, modified to be able to couple the hadronic
states to any reasonable Yang-Mills order parameter.

Finally the present approach has also been extended to strongly
interacting theories with fermions in the fundamental or adjoint
representation of the gauge group. Within our framework we were
then able to understand the relation between confinement and
chiral symmetry breaking, and to make a number of predictions
which can be tested via first principle lattice simulations
\cite{Mocsy:2003qw}.

It is now clear, that the critical behavior defined and governed
by the order parameter can be studied via the singlet fields.


\end{document}